\documentclass[11pt]{article}
\usepackage{jcappub}

\graphicspath{{figures/}}
\usepackage{lipsum}
\usepackage{physics}
\usepackage{multirow}
\usepackage{xspace}
\usepackage{fontawesome5}
\usepackage{xcolor}
\usepackage{wrapfig}
\usepackage[figuresright]{rotating} 
\usepackage[version=4]{mhchem}
\usepackage[hang,flushmargin]{footmisc}

\makeatletter

\newcommand{\checknextarg}{\@ifnextchar\bgroup{\gobblenextarg}{}}
\newcommand{\gobblenextarg}[1]{\,\mathrm{#1}\@ifnextchar\bgroup{\gobblenextarg}{}}
\makeatother

\title{An emulator for the ionizing photon mean free path in ultra-high resolution simulations: the implications of mean free path measurements for the reionization history}

\author[a]{Hurum Maksora Tohfa}
\author[b]{Christopher Cain}
\author[a]{Matthew McQuinn}
\author[c]{Anson D'Aloisio}

\affiliation[a]{Department of Astronomy, University of Washington, Seattle, WA, 98195}
\affiliation[b]{School of Earth and Space exploration, Arizona State University, Tempe, AZ 85281}
\affiliation[c]{Department of Physics \& Astronomy, University of California Riverside, Riverside, CA 92521}

\emailAdd{htohfa@uw.edu}

\abstract{Measurements of the mean free path of ionizing photons from high-redshift quasar spectra at $z \sim 5$-$6$ constrain the reionization history, but interpreting them requires modeling the kiloparsec-scale clumping that large-volume reionization simulations cannot resolve. We present a deep learning emulator for the mean free path (MFP) trained on high-resolution cosmological radiative transfer simulations of ionization fronts sweeping through small 2 comoving Mpc/h volumes. Using a residual multi-layer perceptron neural network, we predict the MFP at a given redshift as a function of the reionization redshift, photoionization rate, wavelength, and box-scale density, achieving a median relative error of 1.3\% across nearly four orders of magnitude in MFP. Integrating its predictions over box-scale overdensity and an extended reionization history allows the emulator to predict the global MFP.  We apply the emulator to extended reionization histories constrained by observed photoionization rates, finding that models prefer late reionization with substantial neutral fractions persisting at $z \lesssim 6$. Fitting a parametric ionization history yields a midpoint of reionization of $z_{\rm re} = 6.58\pm 1.2$ for reionization durations consistent with Planck and kinetic Sunyaev-Zeldovich constraints, and the universe being $10\%$ neutral still at $z < 5.8 ~(6.3)$ at 1~(2)$\sigma$. Global ionizing emissivity inferences using measurements of the photoionization rate and MFP plus our emulator, which avoids common power-law assumptions, suggest a factor of $2-3$ decline between $z = 6$ and $4.8$, in agreement with previous studies. Our method provides an efficient (and more converged) alternative to large-volume radiative-hydrodynamic simulations of reionization for interpreting MFP measurements, and can also serve as a subgrid prescription for the ionizing opacity within such simulations.}


\begin{document}
\maketitle

\section{Introduction}
\label{sec:intro}

Cosmic reionization, the epoch when ionizing radiation from the first stars and quasars ionized the neutral hydrogen in the intergalactic medium (IGM), was the last significant phase transition for the cosmic baryons \cite{1965ApJ...142.1633G, McQuinn2016}. Over the last two decades, observations of the Ly$\alpha$ forest of high-redshift quasars have begun to reveal the details of reionization's last stages, and have constrained its endpoint to $5 < z < 6$~\cite{Fan2006, Becker2015, Kulkarni2019, Keating2019, Nasir2020, Qin2024, Zhu2024, 2024MNRAS.533.1525B}.  %This late reionization may be consistent with the observed properties of galaxies~\cite{Bosman2024, Kakiichi2025, Kashino2025}.
A recent focus, which has corroborated these forest constraints, is measurements of the HI ionizing photon mean free path (MFP), which show rapid evolution between $z = 5$ and $6$ as would be expected as the universe becomes ionized~\cite{Becker2021, Zhu2023, Gaikwad2023, 2024MNRAS.533..676S}.  %However, extracting the underlying ionizing emissivity of galaxies as well as reionization history from the MFP observations requires accurate modeling of the intergalactic medium.  %The mean free path is shaped by the reionization history of the universe and the dynamics of small-scale structure formation in the IGM.

%The MFP quantifies the typical distance ionizing photons travel before being absorbed and is directly observable through continuum absorption in high-redshift quasar spectra up to $z \sim 6$~\cite{Becker2021}.
  Modeling the MFP will allow us to compare with its observations and constrain the reionization history. Additionally, in combination with measurements of the photoionization rate, the MFP can be used to calculate the global ionizing production rate, which has bearing on the number of ionizing photons required to complete reionization as well as maintain an ionized IGM~\cite{Pawlik2009, Davies2021b, Cain2021}. A final motivation is that MFP models may allow us to better simulate reionization and the post-reionization IGM, as the MFP during reionization shapes both its timing and spatial morphology~\cite{Furlanetto2005, McQuinn2007, Alvarez2012, Sobacchi2014, Cain2023}, as well as large-scale fluctuations in the ionizing background after reionization~\cite{McQuinn2011, Davies2016, 2018MNRAS.473..560D}.

Modeling the MFP accurately in simulations is challenging because of the highly multi-scale nature of the problem.  Capturing large-scale fluctuations in density, ionizing background,  etc. in the IGM requires boxes on the order of $\sim 100$ Mpc~\cite{Barkana2004, 2004ApJ...613....1F, McQuinn2007, Iliev2014}, while resolving the Jeans scale of cold, pre-ionized gas requires resolving $\sim$ kpc scales~\cite{Emberson2013,Park2016, 2020ApJ...898..149D, Mao2020, Cain2021}.  Capturing all these scales self-consistently in the entire IGM is computationally intractable, as it would require over five orders-of-magnitude in spatial resolution.  One approach to circumvent this problem, which we take in this work, is to simulate many small regions of the IGM (with high resolution) that sample a representative parameter of large-scale environments.

In this work, we address this challenge by training a deep learning emulator on a suite of high-resolution radiation-hydrodynamic simulations. Our simulations resolve scales down to 2\,$h^{-1}$\,kpc, sufficient to capture the gas clumping and self-shielding that shape the MFP even shortly after a region is reionized. The emulator learns to predict the MFP as a function of redshift, ionizing photon energy, photoionization rate, reionization redshift, and local overdensity. Once trained, it evaluates in milliseconds what would otherwise require hundreds of CPU-hours of simulation time, enabling rapid exploration of the reionization parameter space. 

We apply our emulator to constrain the reionization history from observed MFP measurements at $z > 4.5$. By combining the emulator predictions with measured photoionization rates, we test different reionization scenarios and find that the data prefer late reionization completing at $z \lesssim 6$. We also use the emulator to calculate the ionizing emissivity required to sustain the observed photoionization rates without assuming the frequency-dependence of the opacity is a power-law, finding more gradual evolution around $z = 6$ than previous studies. Our approach provides an efficient framework for interpreting MFP observations without requiring large-volume simulations that struggle to resolve small-scale IGM physics.  It could even be used to provide a subgrid prescription for the ionizing opacity in such simulations.

This paper is organized as follows: in \S\ref{sec:simulations}, we describe our simulations and the deep learning methodology used to construct the MFP emulator, as well as the numerical setup for our analysis to compare our emulator's performance against previous observations and the physical constraints we derive using our emulator. We present our emulator's performance and use it to constrain reionization parameters from observational MFP measurements, and we show our emissivity calculation in section 3, and conclude in section 4.  Appendix A investigates the error of our approximation of ignoring the photoionization rate, and Appendix B shows how to generalize our calculation to include correlations between the emulator variables.  Throughout, we assume the following cosmological parameters: $\Omega_m = 0.305$, $\Omega_{\Lambda} = 1 - \Omega_m$, $\Omega_b = 0.048$, $h = 0.68$, $n_s = 0.9667$ and $\sigma_8 = 0.82$, consistent with~\cite{Planck2018} results. Distances are in comoving units (which we denote as `cMpc' or `ckpc'), unless otherwise specified.

\section{High-Resolution Reionization Simulations}
\label{sec:simulations}

\begin{figure*}
\centering
\includegraphics[width=.8\textwidth]{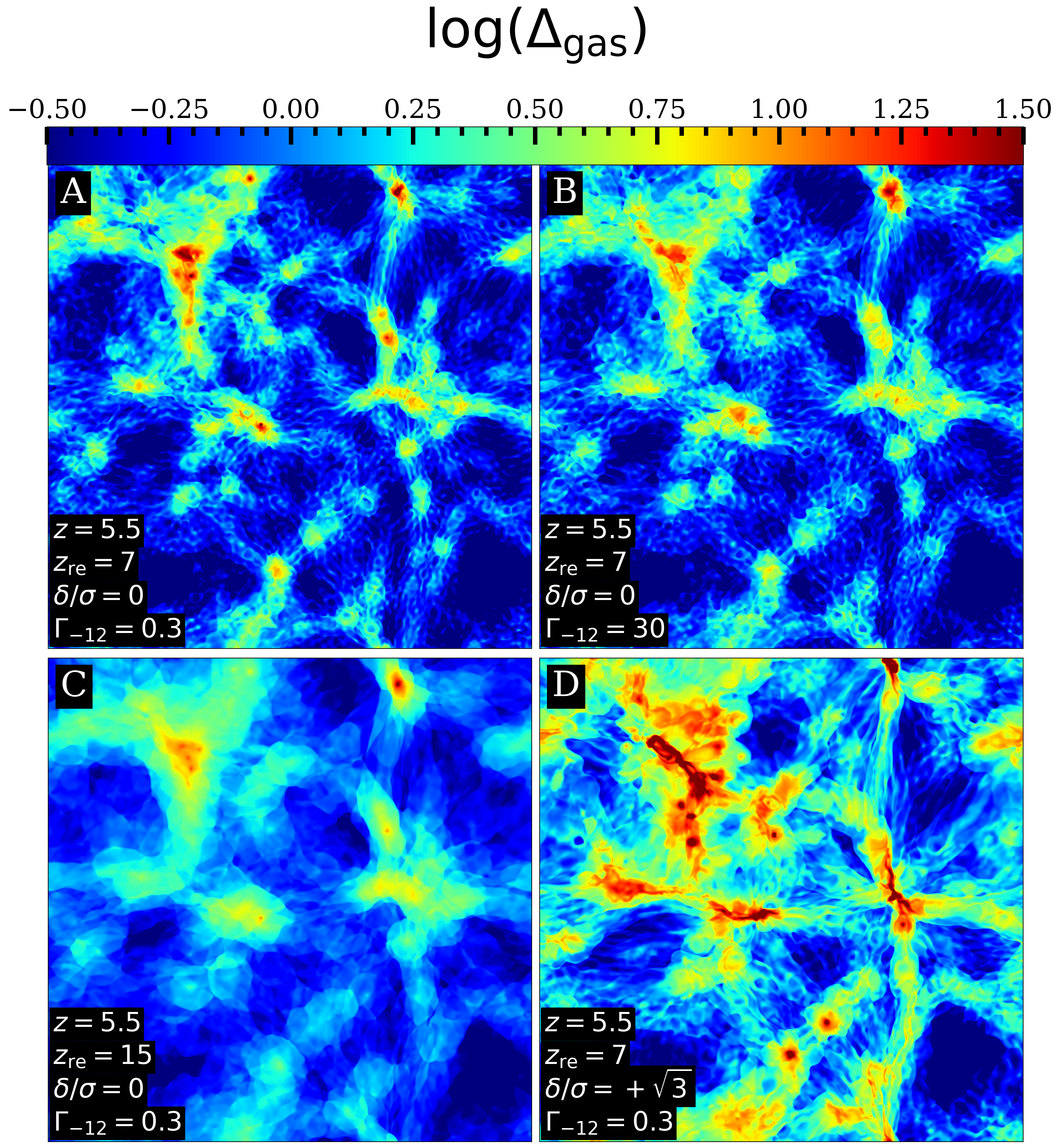}
\caption{Gas density distribution from our simulation suite showing the diversity of IGM environments at $z=5.5$. Each panel shows a 2D slice through a $2\;h^{-1}$ cMpc simulation volume with different combinations of redshift ($z$), reionization redshift ($z_{\rm re}$), photoionization rate ($\Gamma_{-12}$), and box-scale overdensity ($\delta/\sigma$). The colorbar highlights gas densities that range from voids (blue) to dense filaments and minihalos (red/orange).}
\label{fig:sim_environments}
\end{figure*}
Our analysis is based on a comprehensive suite of high-resolution radiation-hydrodynamic simulations that systematically explore the parameter space of reionization environments.  Here, we briefly describe the features of these simulations, and refer the reader to~\cite{2020ApJ...898..149D, Cain2024} for further details.  These simulations employ a modified version of the RadHydro code of \cite{Trac2004,Trac2007}, which solves the radiative transfer (RT) equation using ray tracing fully coupled to gas dynamics on a uniform Eulerian grid. Dark matter is treated by following $N$-body particles with gravity calculated on a particle-mesh. The code uses a reduced speed-of-light approximation to accelerate radiative transfer calculations as described in~\cite{Cain2024} (see also~\cite{2020ApJ...898..149D}). 

All simulations are performed in $L_{\text{box}} = 2\,h^{-1}$ cMpc boxes with $N = 1024^3$ dark matter particles, gas cells, and RT cells with periodic boundary conditions, achieving a spatial resolution of $\Delta x = 2\,h^{-1}$ ckpc. This resolution is sufficient to resolve gas clumping near the Jeans scale and capture the hydrodynamic response of minihalos and filamentary structures to reionization heating \cite{Gnedin2000, Shapiro2004}. While our small box size means we cannot capture density fluctuations on scales larger than $\sim 1$ cMpc, we account for the effects of these large-scale modes using a DC mode approach \cite{Gnedin2011} combined with Gauss-Hermite quadrature. Following \cite{2020ApJ...898..149D}, we run simulations at three box-scale overdensities ($\delta/\sigma = 0, \pm\sqrt{3}$, where $\delta$ is the linearly extrapolated density contrast and $\sigma$ is its standard deviation in our background cosmology on the box scale) corresponding to the roots of the third-order Hermite polynomial. We then compute averages over the cosmic density distribution using three-point Gauss-Hermite quadrature\footnote{This method approximates the integral over all large-scale density fluctuations as a weighted sum over our three simulations, with the $\delta/\sigma = 0$ simulation receiving weight $2\sqrt{\pi}/3$ and the overdense/underdense simulations each weighted by $\sqrt{\pi}/6$. To the extent the averaged quantity can be described by a $\leq 5$ order polynomial, this approach exactly captures the contribution of rare overdense and underdense regions to quantities like the mean free path and recombination rate.}.  The simulations are initialized at $z = 300$ using second-order Lagrangian perturbation theory, and they are evolved to $z = 4$ to capture the post-reionization thermal and density evolution of the IGM.

Following a similar approach to \cite{Cain2024} and \cite{2020ApJ...898..149D}, we model the ionization of IGM gas by external sources using a simplified approach that enables us to label each simulation by its time of ionization. The computational domain is divided into $N_{\rm dom} = 16$ RT domains\footnote{Except in the case of simulations with the lowest ionizing background, in which case we use $N_{\rm dom} = 32$.  }, and plane-parallel rays are cast uniformly from the boundaries of these domains.  The radiation turns on everywhere at a redshift $z_{\rm re}$.  As described in~\cite{Cain2024}, we freeze the dynamical evolution of the gas and the flow of cosmic time at $z_{\rm re}$ long enough for ionization fronts to cross the domains, which has the effect of making reionization happen everywhere at essentially the same redshift and allows us the convenience of categorizing our simulations by $z_{\rm re}$.  The simulations adopt a power-law ionizing spectrum with specific intensity $J_\nu \propto \nu^{-1.5}$ between 1 and 4 Rydberg, distributed across 5 frequency bins \cite{2020ApJ...898..149D}, although this spectrum has little impact on the results.  We later combine the results of these instantaneous-reionization simulations in a manner that allows us to approximate a global reionization history.

Our simulation suite systematically samples the parameter space of reionization environments by varying three key parameters. We use 126 simulations with reionization redshifts $z_{\rm re} = \{5, 6, 7, 8, 9, 12, 15\}$ to span from very late reionization to early reionization in overdense regions. The hydrogen photoionization rate is parameterized by $\Gamma_{-12}$ in units of $10^{-12}$ s$^{-1}$, with values $\Gamma_{-12} = \{0.03, 0.1, 0.3, 1.0, 3.0, 30\}$, encompassing the range of mean ionizing background strengths expected during reionization (as well as much higher values). 

Figure~\ref{fig:sim_environments} shows gas density fields at $z = 5.5$ from our simulation suite, illustrating the impact of reionization redshift ($z_{\rm re}$), photoionization rate ($\Gamma_{-12}$), and box density ($\delta/\sigma$) on IGM structure. All panels show the same cosmic web configuration viewed at identical redshifts but with varying box-scale parameters. Panel A shows a late reionization scenario ($z_{\rm re} = 7$) with low photoionization rate ($\Gamma_{-12} = 0.3$), exhibiting well-defined filamentary structure with deep voids. Panel B demonstrates that a higher photoionization rate ($\Gamma_{-12} = 30$) at the same reionization redshift smooths the density field, as enhanced photoheating suppresses high-density structure. Panel C illustrates an earlier reionization ($z_{\rm re} = 15$) where the IGM has experienced prolonged heating, resulting in significantly smoother density distributions. Panel D shows an overdense region of the universe ($\delta/\sigma = +\sqrt{3}$), where the enhanced large-scale density produces more pronounced structure like denser filaments and more massive halos compared to the mean-density panels.

We calculate mean free path directly from our simulations using the approach of \cite{2015MNRAS.453.2943C}. This approach defines the MFP as the average distance a photon travels weighted by the transmitted flux:
\begin{equation}
\label{eq:mfp}
\lambda_{\rm mfp} = \left\langle \frac{\int x \, df}{\int df} \right\rangle,
\end{equation}
where $f(x) = \exp(-\tau(x))$ is the transmitted ionizing flux at path length $x$ and $\langle ... \rangle$ denotes an average over many lines of sight. Assuming the initial flux at $x = 0$ is normalized to unity, and each sightline is integrated until $f \sim 0$, we have $\int df = 1$, and Eq.~\ref{eq:mfp} becomes
\begin{equation}
    \label{eq:mfp2}
    \lambda_{\rm mfp} = \left\langle \int x df \right\rangle  = \left \langle \sum_{i = 0}^{N} x_i e^{-\tau(x_i)} d\tau_i  \right \rangle,
\end{equation}
where $N$ is chosen large enough that $f(x_N) < 10^{-4}$ and $d\tau_i = n_{\rm HI}^{i} \sigma_{\rm HI} \delta x_i$, where $n_{\rm HI}^{i}$ is the HI number density in cell $i$ and $\sigma_{\rm HI}$ is the HI ionization cross-section at a given frequency.  

We evaluate Eq. 2.1 for 10,000 random sight lines through the simulation volume, which we find is sufficient to converge on the global MFP in the box. We compute the MFP at six energy bins corresponding to 13.6 eV, 14.48 eV, 16.70 eV, 20.05 eV, 25.50 eV, and 39.50 eV to capture the wavelength-dependent opacity. We have also implemented two alternative MFP estimators: a flux-based definition that relates the MFP to the total absorption rate in the simulation (see Appendix C of~\cite{Cain2023}), and a segment-based definition that averages over transmitted flux through short path lengths \cite{Emberson2013, 2020ApJ...898..149D}. These three methods generally agree within a few percent across our parameter space. %This approach directly uses the simulation's neutral hydrogen distribution without requiring parameterized models or assumptions about the functional form of absorption. We focus our analysis on epochs soon after or near the end of hydrogen reionization.

\section{Numerical frameworks and Results}

\subsection{Neural Network Emulator for MFP Prediction}
\label{sec:nn_emulator}

We construct a deep learning emulator to predict the mean free path directly from redshift (z), reionization redshift ($z_{re}$), photoionization rate ($\Gamma_{-12}$), mean density ($\delta/ \sigma$), and energy of ionizing photons. The emulator is trained on MFP values computed at redshifts $4 \le z \le z_{re}$ across the six energy bins described in Section 2. Our emulator uses a residual multi-layer perceptron trained on log-transformed MFP values. We train on $\log_{10}(\lambda_{\rm mfp})$ rather than MFP directly because the MFP spans nearly four orders of magnitude across our parameter space; this compresses the dynamic range and ensures that percentage errors are weighted more uniformly across all MFP values.

We use Huber loss 
 \cite{Huber1964} rather than mean squared error to reduce sensitivity to occasional outliers in the training data, which can arise from stochastic variations in the small simulation volumes. More importantly, it reduces the influence of outliers on training. With standard mean squared error loss in linear space, a few large errors dominate because squaring amplifies them, biasing the network toward extreme cases. The log transformation treats relative errors uniformly. We use Huber loss instead of MSE as it assigns weight quadratically for small residuals but linearly for large ones, making it less sensitive to outliers while maintaining smooth gradients.

Our architecture uses residual connections \cite{he2015deepresiduallearningimage}. Each residual block contains two fully connected layers with layer normalization \cite{Ba2016} and Mish activations \cite{Misra2019}, plus a skip connection that adds the input directly to the output. This lets the network learn residual functions around the identity rather than completely new representations at each layer, facilitating training and avoiding vanishing gradients \cite{Hochreiter1991, Bengio1994}. The complete architecture has an initial layer mapping the five inputs to 128 hidden dimensions, four residual blocks, and a final layer that produces $\log_{10}(\lambda_{\rm mfp})$. We use 0.1 dropout \cite{Srivastava2014} between layers to prevent overfitting.

To improve the network's sensitivity to $z_{\rm re}$, we include an additional term in the training loss that encourages the network to produce more varied predictions when presented with different reionization redshifts.  The additional term has an adjustable weight (set to 2.5) so it guides the network toward better $z_{\rm re}$ sensitivity without compromising its overall accuracy at predicting MFP. By construction, the emulator only returns predictions for $z \leq z_{\rm re}$ (as the mean free path is essentially zero otherwise).

We normalize inputs using scikit-learn's RobustScaler \cite{Pedregosa2011}, which independently scales each input parameter based on its median and interquartile range rather than mean and standard deviation. This makes the scaling less sensitive to extreme values in the training data. We apply the same scaling approach to the log-transformed targets. Training uses the AdamW optimizer \cite{Loshchilov2019} with learning rate 0.005, weight decay 0.001, and a cosine annealing schedule with warm restarts \cite{Loshchilov2017}. We train on batches of 64 samples for up to 5000 epochs with early stopping \cite{Prechelt1998} (patience 450 epochs). We split the entire suite of data into 80\% training, 10\% validation, and 10\% test. The validation set monitors training and selects the best model; all reported metrics are on the held-out test set after converting predictions back to linear space.

\begin{figure*}
\centering
\includegraphics[width=\textwidth]{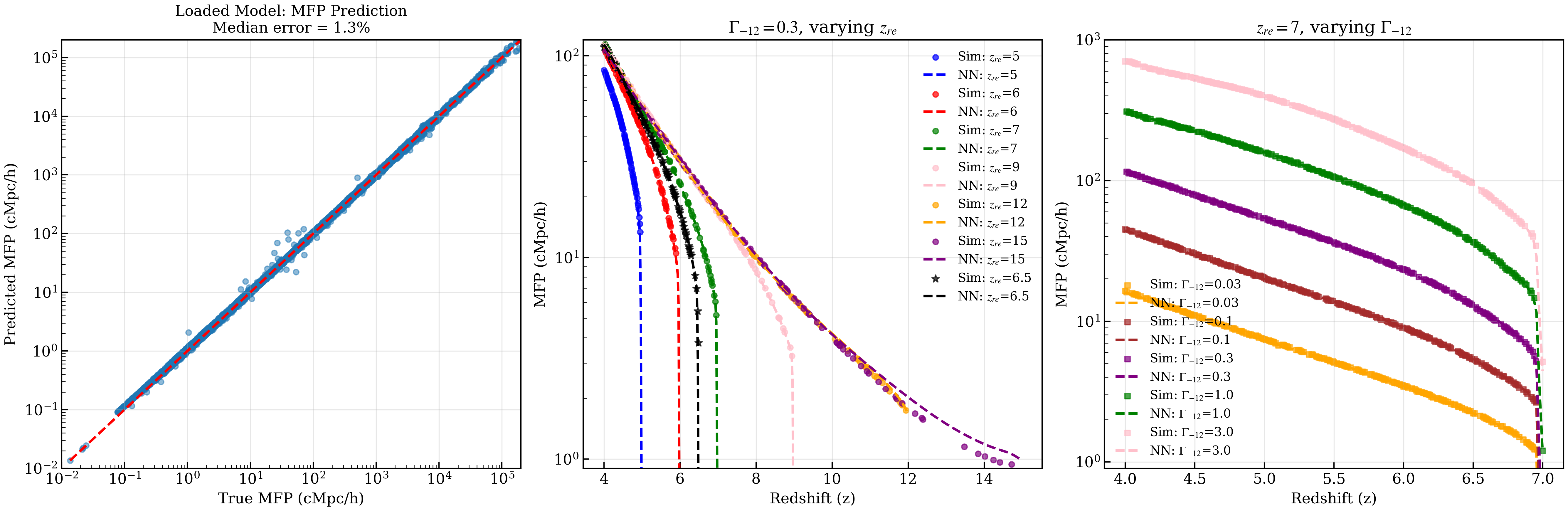}
\caption{Neural network emulator performance and sensitivity to reionization parameters. {\bf Left:} Predicted versus true MFP on the held-out test set, showing excellent agreement with a median relative error of 1.3\%. The red dashed line indicates perfect agreement. {\bf Middle:} MFP evolution as a function of redshift for fixed $\Gamma_{-12} = 0.3$ and varying reionization redshift $z_{\rm re} = \{5, 6, 6.5, 7, 9, 12, 15\}$. Points show direct simulation measurements, while dashed lines show emulator predictions. The black star shows a validation simulation at $z_{\rm re} = 6.5$ not included in the training suite, which the emulator predicts with 1.7\% error at 13.6 eV. {\bf Right:} MFP evolution for fixed $z_{\rm re} = 7$ and varying photoionization rate $\Gamma_{-12} = \{0.03, 0.1, 0.3, 1.0, 3.0\}$. The emulator accurately captures both the amplitude and redshift evolution of MFP across the full parameter space, demonstrating its sensitivity to the physical parameters.}
\label{fig:nn_mfp_performance}
\end{figure*}

Figure~\ref{fig:nn_mfp_performance} summarizes the emulator's performance. The left panel shows predicted versus true MFP on the held-out test set, demonstrating excellent agreement across the full dynamic range with a median relative error of 1.3\% and mean relative error of 2.0\%. Predictions track true values closely across nearly four orders of magnitude in MFP.

The middle and right panels demonstrate the emulator's sensitivity to the key reionization parameters. The middle panel shows MFP evolution for fixed $\Gamma_{-12} = 0.3$ while varying $z_{\rm re}$. Earlier reionization (higher $z_{\rm re}$, purple and yellow curves) produces systematically higher MFP at fixed redshift, as the IGM has had more time to smooth out after photoheating. The emulator predictions (dashed lines) accurately track the simulation measurements (points) across all $z_{\rm re}$ values. As a validation test, we ran an additional simulation at $z_{\rm re} = 6.5$, $\Gamma_{-12} = 0.3$, and mean density that was not included in our training suite, for which the nearest reionization redshifts are $z_{\rm re} = 6$ and  $z_{\rm re} = 7$. The emulator predicts this case with 1.7\% error at 13.6 eV (black star), with errors of 1-3\% across all photon energies. The right panel shows the complementary effect: for fixed $z_{\rm re} = 7$, higher photoionization rates (pink curve with $\Gamma_{-12} = 3.0$) suppress self-shielding and increase MFP, while lower rates (orange curve with $\Gamma_{-12} = 0.03$) allow more widespread neutral hydrogen. The emulator captures both effects accurately. This sensitivity to both $z_{\rm re}$ and $\Gamma_{-12}$ is essential for constraining reionization history from MFP observations.

The 1.3\% median error is sufficiently small for robust constraints on reionization parameters from observational data. The emulator runs in milliseconds on standard hardware, allowing rapid exploration of the full parameter space for inference. This computational efficiency, combined with the physical accuracy inherited from our high-resolution simulations, makes the emulator well-suited for the parameter constraints we present in subsequent sections.

\subsection{Simplified Parameter Constraints from Mean Free Path Observations}
\label{sec:parameter_constraints}

\begin{figure*}
\centering
\includegraphics[width=\textwidth]{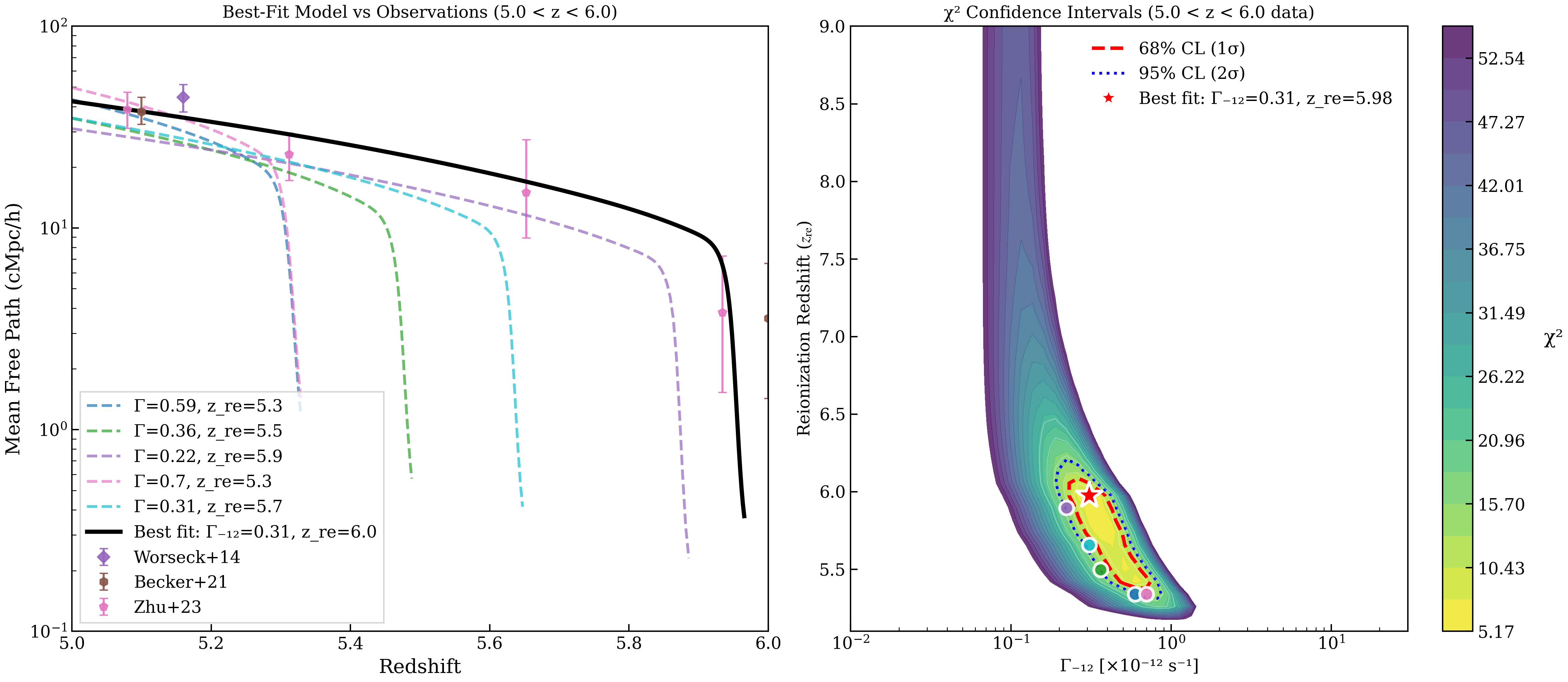}
\caption{Grid search results constraining $\Gamma_{-12}$ and $z_{\rm re}$ from MFP observations at $z > 5.0$, assuming instantaneous reionization. \textbf{Left:} Best-fit model (black line) compared to observations from \cite{Worseck2014} (purple diamonds), \cite{Becker2021} (brown squares), and \cite{Zhu2023} (pink triangles). Several other parameter combinations are shown as dashed lines. The sharp drop in the best-fit model at $z \approx 5.9$ is unphysical for a global average and indicates that dynamical relaxation alone cannot explain the rapid MFP evolution. \textbf{Right:} $\chi^2$ landscape showing the best-fit at $\Gamma_{-12} = 0.31$\,s$^{-1}$ and $z_{\rm re} = 5.98$ (red star). Dashed red and dotted blue contours show 68\% and 95\% confidence regions.}
\label{fig:grid_search_results}
\end{figure*}

To validate our MFP constraint framework, we first consider the simplest case where the universe ionizes instantaneously at a single redshift with a fixed photoionization rate. This provides a baseline for comparison before incorporating extended reionization histories in later sections. Despite the simplicity of the model, it gives some insight into when reionization likely must end to be consistent with MFP measurements. To find the models that are most consistent with measurements, we search over a grid of models specified by $\Gamma_{-12}$ and $z_{\text{re}}$, comparing the emulator's mean free paths with observational measurements compiled from multiple independent studies \cite{Becker2021, Worseck2014, Gaikwad2023, Zhu2023}.\footnote{We exclude MFP measurements from \cite{Gaikwad2023} from this parameter optimization. While their photoionization rate measurements are derived from direct comparison of effective optical depth distributions with observations, their MFP values are obtained through a model-dependent transformation involving their EX-CITE code, which uses assumed relationships between mean free path, local density, and photoionization rate fluctuations. This introduces additional model dependence that makes direct comparison with our emulator predictions less straightforward. In contrast, the MFP measurements from \cite{Becker2021}, \cite{Worseck2014}, and \cite{Zhu2023} are derived more directly from stacking quasar spectra at the Lyman limit (912~\AA). %We note that these stacking measurements do not measure the MFP exactly as we define it in Equation~2, but at $z \gtrsim 5$ the measured quantity is very close to the MFP \cite{2024MNRAS.530.5209R}. 
The high-redshift ($z > 5$) MFP measurements from \cite{Becker2021}, \cite{Worseck2014}, and \cite{Zhu2023}, particularly at $z = 6$, do carry some model dependence related to assumptions about the quasar continuum shape and proximity zone corrections. However, these studies directly measure continuum opacity near the Lyman limit, and \cite{2024MNRAS.530.5209R} demonstrated that this direct approach recovers the true MFP accurately when accounting for model-dependence, with reported uncertainties including marginalization over these effects. In contrast, \cite{Gaikwad2023} infer the MFP indirectly through their EX-CITE model from Ly$\alpha$ forest measurements. While we do compare our best-fit model predictions with the full range of observational constraints including \cite{Gaikwad2023} in subsequent analysis, the model-dependent transformation in their approach makes using their MFP values in the grid search optimization less straightforward.}

Our grid search explores the parameter ranges $\Gamma_{-12} \in [0.03, 3]$ (logarithmic spacing) and $z_{\text{re}} \in [5.0, 15.0]$ (linear spacing) using only our mean density simulations\footnote{The Gaussian quadrature integration over density (Section 2) produces nearly identical constraints in this simplified scenario, so we use only the mean density for computational efficiency. We apply the full density integration in Section 3.4-3.5 for extended reionization histories.}, evaluating $200 \times 200 = 40,000$ parameter combinations. Traditional simulation-based approaches would require running this number of individual high-resolution simulations, which would be prohibitively expensive. Our emulator evaluates all combinations in a few minutes. For each parameter combination $(\Gamma, z_{\text{re}})$, we use the neural network emulator to predict mean free paths and compute a $\chi^2$ statistic:
\begin{equation} 
\chi^2 = \sum_{i} \frac{(\lambda_{\text{obs},i} - \lambda_{\text{pred},i})^2}{\sigma_{\text{obs},i}^2}, 
\end{equation} 
where the summation is over all observational data points.

Figure~\ref{fig:grid_search_results} presents the results of this grid search. The right panel displays the $\chi^2$ landscape, with a well-defined minimum at $\Gamma_{-12} = 0.31 \pm 0.22$ and $z_{\rm re} = 5.98 \pm 0.32$. The confidence contours show the expected degeneracy: earlier reionization compensates for weaker photoionization rates. This occurs because earlier reionization allows more time for the relaxation of the IGM structure owing to pressure imbalances from the heating at ionization. To match the observed low MFP values, models with earlier $z_{\rm re}$ therefore require lower $\Gamma_{-12}$ values. Despite this degeneracy, the minimum is tightly constrained, with $\chi^2$ rising steeply away from the best-fit. The disconnected island of low $\chi^2$ at $\Gamma_{-12} \sim 10$ in Figure~\ref{fig:grid_search_results} corresponds to an unphysical degeneracy where a very high photoionization rate compensates for a very low reionization redshift. So we do not investigate this region further.

The left panel demonstrates both the quality and limitations of the fit. The predicted MFP evolution (black line) tracks observations from the three studies used in the fitting, but exhibits an unphysical sharp drop at $z \approx 5.9$. Several other parameter combinations are shown as dashed lines, and all models that attempt to match the rapid MFP decline between $z = 5.6$ and $6$ exhibit similarly sharp drops. While individual small patches of the IGM can experience such rapid MFP evolution for brief periods just after their local $z_{\rm re}$ as  gas is ionized and then photoevaporated, this behavior is  unrealistic for a global average. The $z = 6$ measurement, combined with the higher MFP values at $z < 5.8$, forces any instantaneous reionization model with constant $\Gamma_{-12}$ into this unphysical regime.

Yet, noting the simplistic nature of the model, the values that come out capture the ethos of many reionization constraints. Our best-fit $z_{\rm re} = 5.98 \pm 0.32$ is just somewhat lower than Planck CMB constraints placing the midpoint of reionization at $z = 7.8 \pm 0.9$~\cite{PlanckXLVII2016}, and consistent with the large number of Ly$\alpha$ forest studies that conclude that reionization is still finishing somewhat below $z=6$ \cite{Kulkarni2019,Nasir2020,Zhu2024}. Lyman-$\alpha$ emitter surveys also suggest that the universe is still significantly neutral at $z = 6.5$ \cite{Mason2018a,Mason2019,Morales2021,Nakane2023,Tang2024b}.  Our constraint on $\Gamma_{-12}$ is also in line with measurements at $z=5$ \cite{2018MNRAS.473..560D,Gaikwad2023,Davies2024}, although these measurements show some evolution that our simplistic model ignores. 

In the following section, we use observed $\Gamma_{-12}$ values alongside MFP measurements to further constrain the reionization history, incorporating the patchy nature of reionization and time-evolving photoionization rates rather than assuming instantaneous completion with fixed $\Gamma_{-12}$.

\subsection{Mean Free Path Predictions for Patchy Reionization}
\label{sec:patchy_reionization}

The instantaneous reionization analysis in the previous section is a simplification. Reionization occurred over hundreds of millions of years and is spatially inhomogeneous, meaning that different regions completed reionization at different times. We now test how this patchiness affects MFP predictions by incorporating realistic reionization histories.

\begin{figure*}
\centering
\includegraphics[width=\textwidth]{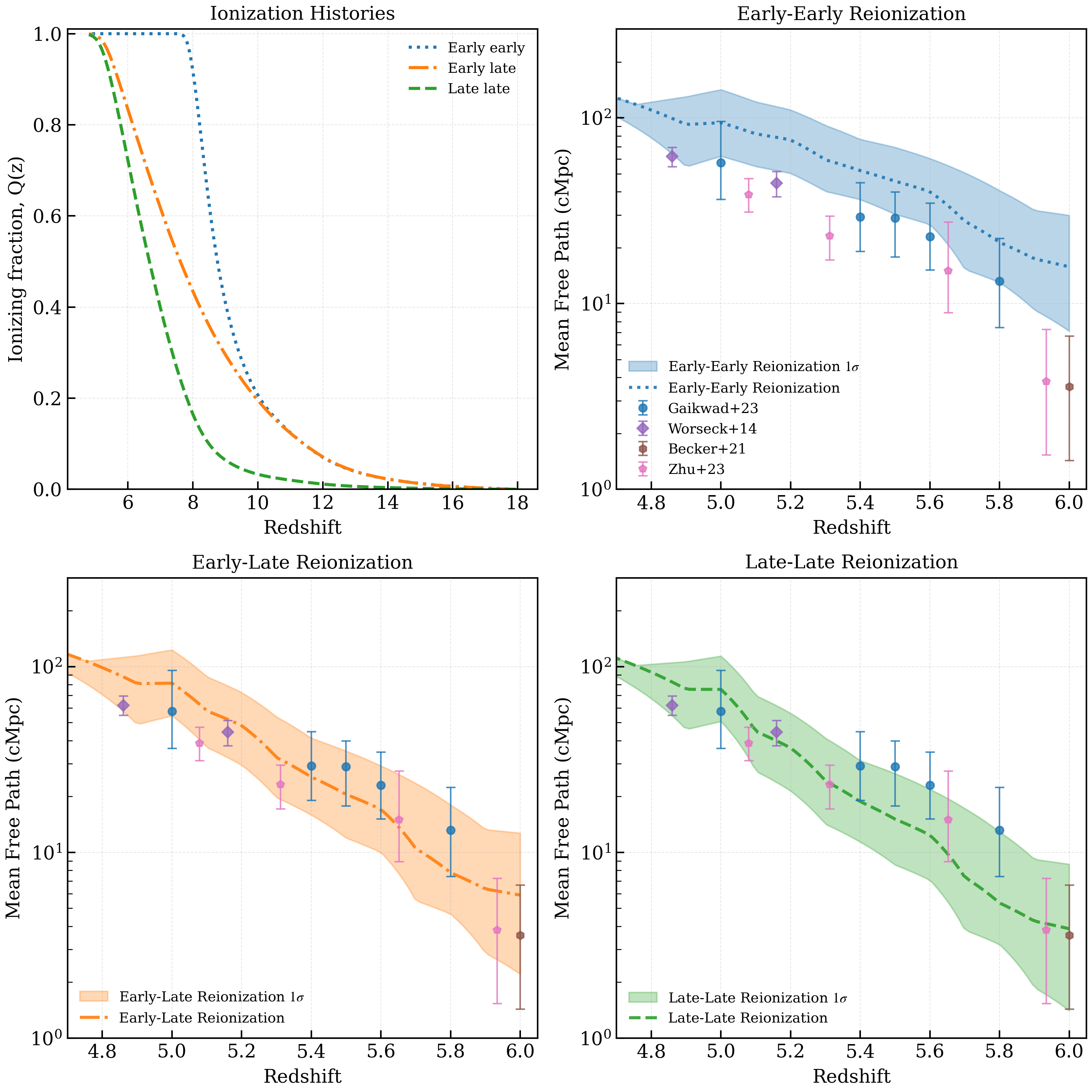}
\caption{Mean free path predictions for three reionization histories compared to observations at $z > 4.5$. Each panel shows a different reionization scenario: early-early (top right), early-late (bottom left), and late-late (bottom right). Colored points show observations from multiple studies, and solid lines show emulator predictions using measured photoionization rates from the same studies. Shaded bands show uncertainties propagated from $\Gamma_{-12}$ measurement errors. All three models provide reasonable agreement with observations when using measured $\Gamma_{-12}$ values.}
\label{fig:gamma-reionization}
\end{figure*}

To account for the extended nature of reionization, we integrate our opacity, $\kappa = 1/\lambda_{\rm mfp}$ predictions over the reionization history:
\begin{equation}
\langle \kappa \rangle = \int_0^\infty dz_{\rm re} \, P(z_{\rm re}) \, \kappa(z_{\rm re}),
\label{eqn:opacity}
\end{equation}

where $P(z_{\rm re})\, dz_{\rm re}$ is the volume-weighted probability that a randomly chosen patch of the IGM was reionized between $z_{\rm re}$ and $z_{\rm re} + dz_{\rm re}$. Because the volume-averaged ionized fraction $Q(z)$ is, by definition, the cumulative fraction of the IGM that has been reionized by redshift $z$, the probability density of the local reionization redshift is
\begin{equation}
P(z_{\rm re}) = -\left.\frac{dQ}{dz}\right|_{z = z_{\rm re}},
\label{eq:Pzre}
\end{equation}
where the minus sign accounts for $Q$ decreasing with increasing redshift. This is the same volume-weighted distribution used in semi-numerical and parametric models of patchy reionization~\cite{2004ApJ...613....1F, Battaglia2013}, properly accounting for the distribution of local ionization times. Regions that ionize early contribute opacity through long-relaxed gas, while late-ionizing regions contribute opacity from gas that has had less time to dynamically respond to photoheating.

Our 2 cMpc/$h$ simulations do not capture density fluctuations on scales larger than the box. We account for these large-scale modes using separate-universe simulations at different box-scale overdensities \cite{2015MNRAS.448L..11W} using DC modes. The volume-averaged opacity becomes \cite{2020ApJ...898..149D}:
\begin{equation}
\langle \kappa \rangle = \eta \int_{-\infty}^{\infty} d\delta \, \frac{\kappa(\delta)}{1 + \delta_{\rm NL}(\delta)} \, P_V(\delta),
\label{eqn:densityint}
\end{equation}
where $P_V(\delta)$ is the volume-weighted probability distribution of the linear overdensity, the factor $(1 + \delta_{\rm NL})^{-1}$ accounts for the volume occupied by Lagrangian elements after gravitational collapse, and $\eta$ is a normalization factor. For a Gaussian linear overdensity field, $P_V(\delta) = (2\pi\sigma_z^2)^{-1/2} \exp(-\delta^2/[2\sigma_z^2])$, where $\sigma_z$ is the variance of the linear overdensity smoothed on the box scale. We evaluate this integral using three-point Gauss-Hermite quadrature as described in Section 2, computing the emulator predictions at our three simulation densities $\delta/\sigma = \{-\sqrt{3}, 0, +\sqrt{3}\}$ and combining them with the appropriate quadrature weights. This density integration decreases the predicted MFP by $10\%$  compared to the mean density.

The above integrals over $z_{\rm re}$ and $\delta / \sigma$ (Eqns.~\ref{eqn:opacity},~\ref{eqn:densityint}) can be generalized to a three dimensional integral that also includes $\Gamma_{-12}$.  This would allow including small correlations between all the variables, as well as spatial fluctuations in $\Gamma_{-12}$.  Appendix~\ref{ap:correlations} discusses this generalization and why we do not expect it will be important for our approach in the paper, which ignores these effects.

Unlike our earlier approach which also fit for a redshift-independent $\Gamma_{-12}$, here we use measurements of the photoionization rates (including uncertainties) as input. When multiple measurements exist at the same redshift, we adopt the most recent determination. The photoionization rate measurements are compiled from \cite{2011MNRAS.412.2543C}, \cite{2011MNRAS.412.1926W}, \cite{2018MNRAS.473..560D}, \cite{Becker2021}, and \cite{Gaikwad2023}.\footnote{For reproducibility, the adopted $\Gamma_{-12}$ values (in units of $10^{-12}$ s$^{-1}$, prioritizing the most recent studies) are: $z=4.8$: $0.58^{+0.08}_{-0.20}$; $z=4.9$: $0.501^{+0.275}_{-0.232}$; $z=5.0$: $0.557^{+0.376}_{-0.218}$; $z=5.1$: $0.508^{+0.324}_{-0.192}$; $z=5.2$: $0.502^{+0.292}_{-0.193}$; $z=5.3$: $0.404^{+0.272}_{-0.147}$; $z=5.4$: $0.372^{+0.217}_{-0.126}$; $z=5.5$: $0.344^{+0.219}_{-0.130}$; $z=5.6$: $0.319^{+0.194}_{-0.120}$; $z=5.7$: $0.224^{+0.223}_{-0.112}$; $z=5.8$: $0.178^{+0.194}_{-0.078}$; $z=5.9$: $0.151^{+0.151}_{-0.079}$; $z=6.0$: $0.145^{+0.157}_{-0.087}$.} Since this specifies $\Gamma_{-12}$, the emulator can then predict the MFP with redshift given a model for $Q(z)$.

We first consider three global ionization histories shown in Figure~\ref{fig:gamma-reionization}: (1) early-early reionization, where reionization begins at $z \approx 12$ and completes by $z \approx 8$; (2) early-late reionization, beginning at $z \approx 12$ but completing as late as $z \approx 6$; and (3) late-late reionization, where both onset and completion are delayed to lower redshifts. For each scenario, we compute $P(z_{\rm re})$ from the global ionizing history using Eq.~\eqref{eq:Pzre}. We propagate the $\Gamma_{-12}$ measurement errors through the emulator to obtain uncertainty bands on the MFP predictions.

Figure~\ref{fig:gamma-reionization} compares the three reionization scenarios against observations from $z = 4.5$ to $6$. When using measured $\Gamma_{-12}$ values and uncertainties as input, the shaded uncertainty bands from $\Gamma_{-12}$ measurement errors overlap the observational data points for the two late reionization models (early-late reionization model in yellow dashed line and shaded region, and late-late model in green dashed line and shaded region) within the $1\sigma$ range and tends to prefer a later reionization history. The early-early reionization model (dotted blue line and blue region) fails to reproduce the observed mean free path evolution, consistently overpredicting it by factors of 2-3 and placing it outside the observational uncertainties.  

One concern is that our emulator is trained on models that do not include evolution in $\Gamma_{-12}$, as the simulations assume constant $\Gamma_{-12}$.  Fortunately, the evolution of $\Gamma_{-12}$ is a secondary effect on the MFP as the photoheating of optically thin gas does not depend on the amplitude of the ionizing background, and the current ionization state is most sensitive to the current value of $\Gamma_{-12}$.  We note that systematic uncertainties from $\Gamma_{-12}$ evolution (discussed in Appendix~\ref{sec:evolving_gamma}, which bounds the effect) could reduce the tension with the early-early scenario somewhat, though it would still remain disfavored at $2\sigma$. %In this scenario, the early-late model will be most favored by the data.

\subsection{Constraining a Parametric Model for the Ionization History}
\label{sec:tanh_constraints}

\begin{figure*}
\centering
\includegraphics[width=\textwidth]{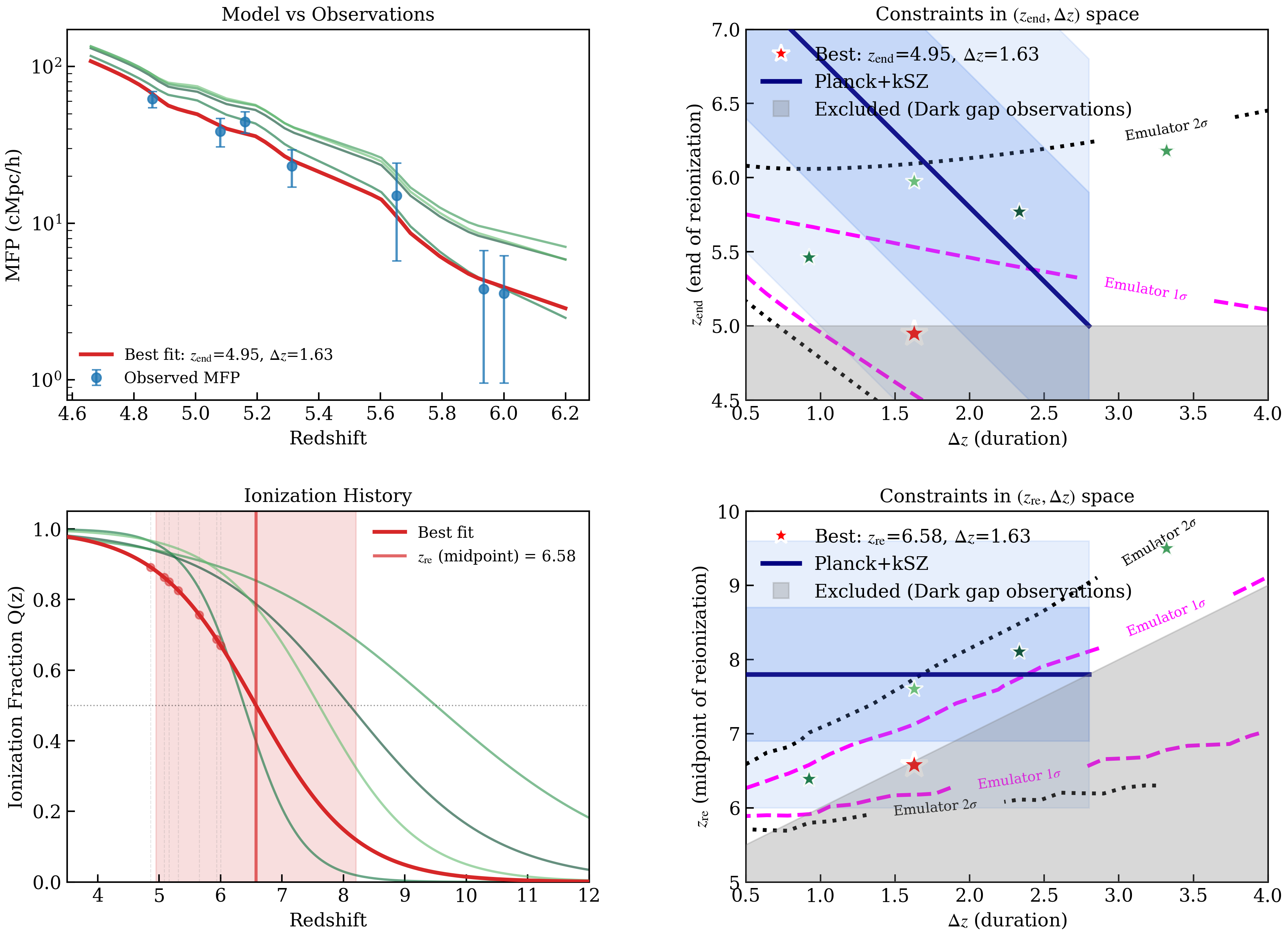}
\caption{Constraints on the ionization history from a tanh model fit to MFP observations. {\it Top Left:} Best-fit model (red line) compared to observed MFP values (blue points with error bars). Sample parameter combinations are shown as thin colored lines, whose parameter values are the corresponding stars in the other panels with the red star marking the best fit.  {\it Top Right:} $\chi^2$ landscape in the $(z_{\rm end}, \Delta z)$ plane.  Dashed magenta and dotted black lines show 68\% and 95\% confidence regions from the MFP fit. The blue band shows Planck+kSZ (using SPT and ACT data) CMB constraints \cite{PlanckXLVII2016} ($z_{\rm re} = 7.8 \pm 0.9$, $\Delta z < 2.8$ at 68\% confidence) with $z_{\rm re} = 7.8$ represented by the navy blue line. The grey shaded region shows parameter space excluded by dark gap observations \cite{Davies_2025}, which constrain $z_{\rm end} > 5$ at $2 \, \sigma$. Colored stars mark sample parameter combinations shown in the other panels. {\bf Bottom Left:} Best-fit ionization history (red line) with sample parameter combinations shown as thin colored lines matching the stars in the right panels. The vertical navy blue line and blue shaded region indicate $z_{\rm re} = 6.58$ (midpoint) and the $\pm\Delta z$ range. Red circles mark the ionization fractions at the observed redshifts. {\bf Bottom Right:} Same $\chi^2$ landscape shown in the $(z_{\rm re}, \Delta z)$ plane for comparison with Planck+kSZ constraints. The grey shaded region again shows the dark gap exclusion. Our MFP-based constraints are consistent with Planck+kSZ at the 1$\sigma$ level.}
\label{fig:tanh_fit}
\end{figure*}

The previous section tested instantaneous reionization scenarios drawn from simulations. We take a complementary approach by fitting a parametric ionization history directly to the MFP observations. We adopt a tanh model for the volume-averaged ionization fraction:
\begin{equation}
Q(z) = \frac{1}{2}\left[1 + \tanh\left(\frac{z_{\rm re} - z}{\Delta z}\right)\right],
\label{eqn:tanh}
\end{equation}
where $z_{\rm re}$ is the midpoint of reionization (when $Q = 0.5$) and $\Delta z$ characterizes its duration. We parametrize the fit in terms of $z_{\rm end}$ (the redshift when reionization ends, defined as $Q(z_{\rm end}) \approx 0.88$) and $\Delta z$, with the relation $z_{\rm re} = z_{\rm end} + \Delta z$. This parametrization directly constrains when reionization completes rather than its midpoint.

We generalize our MFP calculation by ignoring the contribution to the MFP from neutral regions:
\begin{equation}
\lambda_{\rm mfp}(z)^{-1} = Q(z)^{-1} \int_{z}^\infty dz_{\rm re} \, P(z_{\rm re}) \, \kappa(z_{\rm re}),
\end{equation}
The motivation for this generalization is that the patchy structure of reionization results in MFP measurements fitting to a transmission profile blueward of the Lyman limit that $\propto Q(z)\exp(-x/\lambda_{\rm mfp})$. That is, sightlines beginning in neutral regions contribute no transmission—quasars embedded in neutral regions would show complete absorption below 912~\AA\ -- while sightlines through ionized regions contribute the $\exp(-x/\lambda_{\rm mfp})$ profile that our emulator models. Everywhere else, the ionized regions that our emulator is modeling contribute the $\exp(-x/\lambda_{\rm mfp})$ profile.\footnote{Quasar proximity regions may provide another term that is $\propto 1- Q(z)$ with a smaller effective mean free path, rather than zero transmission as would otherwise occur in neutral regions. This does not change our argument.} The key point is that in all realistic models $Q(z)\gtrsim 0.7$ for the $z<6$ mean free path measurements we compare to, so that the inferred MFP should be close to $\lambda_{\rm mfp}$ (particularly because of the exponential dependence of the profile on $\lambda_{\rm mfp}$). Thus, our approximation of the MFP as $\lambda_{\rm mfp}$ will slightly overestimate the mean free path (and hence bias the model toward somewhat higher reionization redshifts), but this overestimate should be small.

For each point in the $(z_{\rm end}, \Delta z)$ parameter space, we compute the reionization-weighted MFP by integrating over $P(z_{\rm re})$ (Eq.~\eqref{eq:Pzre}), as described in Section~\ref{sec:patchy_reionization}. As also described there, we use measurements of $\Gamma_{-12}(z)$ as inputs to our model. We propagate the $\Gamma_{-12}$ uncertainties into the MFP predictions using $d\log\lambda_{\rm mfp}/d\log\Gamma_{-12} \approx 2/3$, which we find is the approximate scaling of MFP with photoionization rate in our models and this scaling was also motivated in \cite{McQuinn2011}. The total error budget thus includes both the observational MFP uncertainties and the contribution from $\Gamma_{-12}$ measurement errors.

We perform a grid search over $z_{\rm end} \in [3.0, 7.0]$ and $\Delta z \in [0.5, 6.0]$, evaluating 1600 parameter combinations. Figure~\ref{fig:tanh_fit} shows the results. The top left panel demonstrates excellent agreement between our best-fit model (red line) and observations. The top right panel displays the $(z_{\rm end}, \Delta z)$ plane with confidence contours at 68\% (dashed magenta) and 95\% (dotted black).  Since $z_{\rm end} = z_{\rm re} - \Delta z$, $z_{end}$ corresponds to when the IGM is 12\% neutral with our tanh parametrization (Eqn.~\ref{eqn:tanh}).  The grey shaded region indicates parameter space excluded by dark gap observations \cite{Davies_2025}, which constrain $z_{\rm end} > 5$ based on measurements of the distribution of dark pixel gaps in the Ly$\alpha$ forest (namely, their 2$\sigma$ lower limit in their $4.95<z<5.15$ redshift bin is $ Q> 0.90$). Much of our $1 \sigma$ region  is also consistent with this constraint. The mild tension we see is not severe as MFP measurements are relatively insensitive to models with extended tails of low neutral fractions at $z < 5$, which can still produce dark gaps. The dark gap constraints are complementary to our MFP-based analysis, as they probe different aspects of the tail end of reionization.

Our MFP-based constraints are also consistent with Planck+kSZ at the 1$\sigma$ level \cite{PlanckXLVII2016}. The $z_{\rm re}$ midpoint is well constrained by Planck measurements of the Thomson scattering optical depth $\tau$, which is primarily sensitive to the integrated electron column density and thus the reionization midpoint. The duration $\Delta z$ is more tightly constrained by measurements of the patchy kinetic Sunyaev-Zeldovich (kSZ) effect, which is sensitive to the patchiness of reionization and thus its duration. Since Planck's signal-to-noise decreases rapidly above $\ell = 2000$ where the kSZ signal peaks, \cite{PlanckXLVII2016} combined Planck data with high-resolution CMB temperature power spectrum measurements from both the Atacama Cosmology Telescope \cite{2014JCAP...04..014D} and South Pole Telescope \cite{2015ApJ...799..177G}, covering multipoles up to $\ell = 13000$. The combined Planck+kSZ constraints favor a reionization midpoint $z_{\rm re} = 7.8 \pm 0.9$ and $\Delta z < 2.8$ at 68\% confidence, shown by the blue band in Figure~\ref{fig:tanh_fit}. The MFP data prefer a slightly later completion of reionization but remain consistent at 1$\sigma$. The bottom panels provide complementary views of the constraints. The bottom right panel shows the same constraints but in the $(z_{\rm re}, \Delta z)$ plane, which is the parameter space where the Planck+kSZ covariance matrix is diagonal with $z_{\rm re} \approx 7.8 \pm 0.9$. The grey shaded region shows the same dark gap exclusion transformed to this parameter space.

The bottom left panel shows the best-fit ionization history (red line) along with sample parameter combinations (thin colored lines) that correspond to the colored stars in the right panels. The colors are matched consistently across all panels to show each model's ionization history, MFP evolution (top left panel), and location in parameter space. The red circles mark the ionization fractions at the observed redshifts, showing that our MFP measurements probe the final stages of reionization when $Q(z) = 0.5$-$0.9$. The vertical red line indicates the midpoint $z_{\rm re} = 6.58$, and the shaded region shows the $\pm \Delta z$ range, demonstrating that reionization begins at high redshift but completes relatively late at $z \lesssim 5$.

The best-fit parameters required by our model imply an extended reionization history, with a neutral fraction of $\approx 30\%$ at $z = 6$. This is higher than some observationally-based inferences \cite{Gaikwad2023,Zhu2024,Qin2024} but consistent with others that find substantial neutral fractions persisting at $z < 6$. A late reionization ending after $z\approx 6$ is strongly preferred by the MFP data. This preference arises from the fast evolution of the MFP between $z = 5.6$ and $z = 6$. In addition to the assumed evolution in $\Gamma_{-12}$, our model requires substantial evolution in the clumpiness of the IGM to explain this behavior, which demands that a significant fraction of the IGM was recently ionized during this redshift range.

Note that since we are not including opacity from neutral islands, our model does not take into account additional opacity evolution that would arise if reionization ended late. Our result shows that MFP evolution assuming a fully ionized IGM indirectly leads to the conclusion that the IGM is partially neutral at $z \approx 6$.  The excellent fit quality when allowing for extended reionization, combined with the 1$\sigma$ consistency with Planck+kSZ constraints, suggests that any self-consistent model that can fully explain the observed MFP evolution must include opacity from neutral islands persisting to $z \lesssim 6$. This even holds up when in our models in Appendix~\ref{sec:evolving_gamma} that account for the maximum effect on the MFP of $\Gamma_{-12}$ evolution.

\subsection{Ionizing Emissivity from the MFP Emulator}

The comoving emissivity is one of our best observables for assessing how many ionizing photons galaxies produce. We connect our MFP predictions to the global ionizing photon budget by calculating the emissivity required to sustain the observed photoionization rates. The volume-averaged ionization rate relates to the spectral emissivity through:
\begin{equation}
\Gamma = (1+z)^2 \int d\nu \, \dot{N}_\gamma(\nu) \sigma(\nu) \lambda(\nu),
\label{eqn:gamma}
\end{equation}
where $\dot{N}_\gamma(\nu)$ is the comoving spectral emissivity (ionizing photons per unit time, comoving volume, and frequency), $\sigma(\nu)$ is the hydrogen photoionization cross-section, and $\lambda(\nu)$ is the comoving frequency-dependent mean free path.\footnote{Equation~(\ref{eqn:gamma}) assumes the number of ionizing photons produced is in direct balance with the number of case-B recombinations. In practice, the ionizing photons travel for some time before being absorbed, during which redshifting and intensity dilution reduces their contribution to the photoionization rate. These effects are largest when the photons travel for an appreciable fraction of the age of the Universe, becoming order unity at $z\sim 2$. They are less important for the high redshifts that are our focus. \cite{Cain2025} finds that these effects lead to equation~(\ref{eqn:gamma}) underpredicting by 20\% the emissivity at the lowest redshifts we consider, $z\approx5$. We ignore this correction here.}

We parameterize the source spectrum as $\dot{N}_\gamma(\nu) = \dot{N}_\gamma \phi_\nu$, where $\phi_\nu = \phi_0 \, \nu^{-(\alpha_s+1)}$ describes the normalized spectral energy distribution. We adopt $\alpha_s = 1.5$ based on models for low-metallicity stellar populations, although models with extremely low metallicities or stellar binaries may indicate values as low as $\alpha_s = 0.5$ \cite{DAaloisio2019}. The normalization condition $\int d\nu \, \phi_\nu = 1$ determines $\phi_0$, and we integrate over the energy range $E = 13.6$--$54.4$\,eV corresponding to the H\,\textsc{i} and He\,\textsc{ii} ionization thresholds.

For each frequency, we compute the wavelength-dependent mean free path $\lambda(\nu)$ using our emulator. We adopt the late-late and early-late reionization history from Section~\ref{sec:patchy_reionization} as they were most favored by the data and integrate over the probability distribution $P(z_{\rm re})$ derived from the ionization fraction evolution. By assuming a reionization model, our emissivity constraints do not consider the uncertainty in histories that are consistent with the MFP measurements, and so the errors are underestimated. However, our emulator is precisely the type of tool that would allow performing this marginalization. We also integrate over box-scale overdensities using three-point Gaussian quadrature at $\delta/\sigma = \{-\sqrt{3}, 0, +\sqrt{3}\}$ to account for large-scale density fluctuations. This approach captures the extended nature of reionization, the inhomogeneous density field, and the wavelength-dependent mean free path without assuming power-law scaling relationships for the opacity.

For comparison, we also calculate $\dot{N}_{\rm ion}$ using the common approximation that assumes power-law scaling for the source spectrum and opacity \cite[e.g.][]{2007MNRAS.382..325B}:
\begin{equation}
\begin{split}
\dot{N}_{\rm ion} &\simeq 10^{51.2} \, \Gamma_{-12} \left(\frac{\alpha_s}{1.5}\right)^{-1} \left(\frac{\alpha_b+3}{3}\right) \\
&\quad \times \left(\frac{\lambda_{\rm mfp}}{40~{\rm Mpc}}\right)^{-1} \left(\frac{1+z}{7}\right)^{-2} \, {\rm s}^{-1} \, {\rm cMpc}^{-3}.
\end{split}
\label{eqn:Nion}
\end{equation}
For calculating $\dot{N}_{\rm ion}$, we interpolate the observed photoionization rates $\Gamma_{-12}$ to match the redshifts of the MFP measurements. We adopt this approach because the $\Gamma_{-12}$ values are more densely sampled (with most recent measurements from \cite{Gaikwad2023}) while the MFP measurements are sparser and come from multiple independent studies \cite{Becker2021, Worseck2014, Zhu2023}, making interpolation of $\Gamma_{-12}$ more reliable than interpolating the MFP.  The spectral index of the background relates to that of the sources via $\alpha_b = \alpha_s - 3(\beta - 1)$, where $\beta$ is the power-law slope of the HI column density distribution ($d\mathcal{N}/dN_{\rm HI}$). Comparing our direct calculation to equation~(\ref{eqn:Nion}) thus indicates the effective slope of $d\mathcal{N}/dN_{\rm HI}$.

\begin{figure}
    \centering
    \includegraphics[width=\columnwidth]{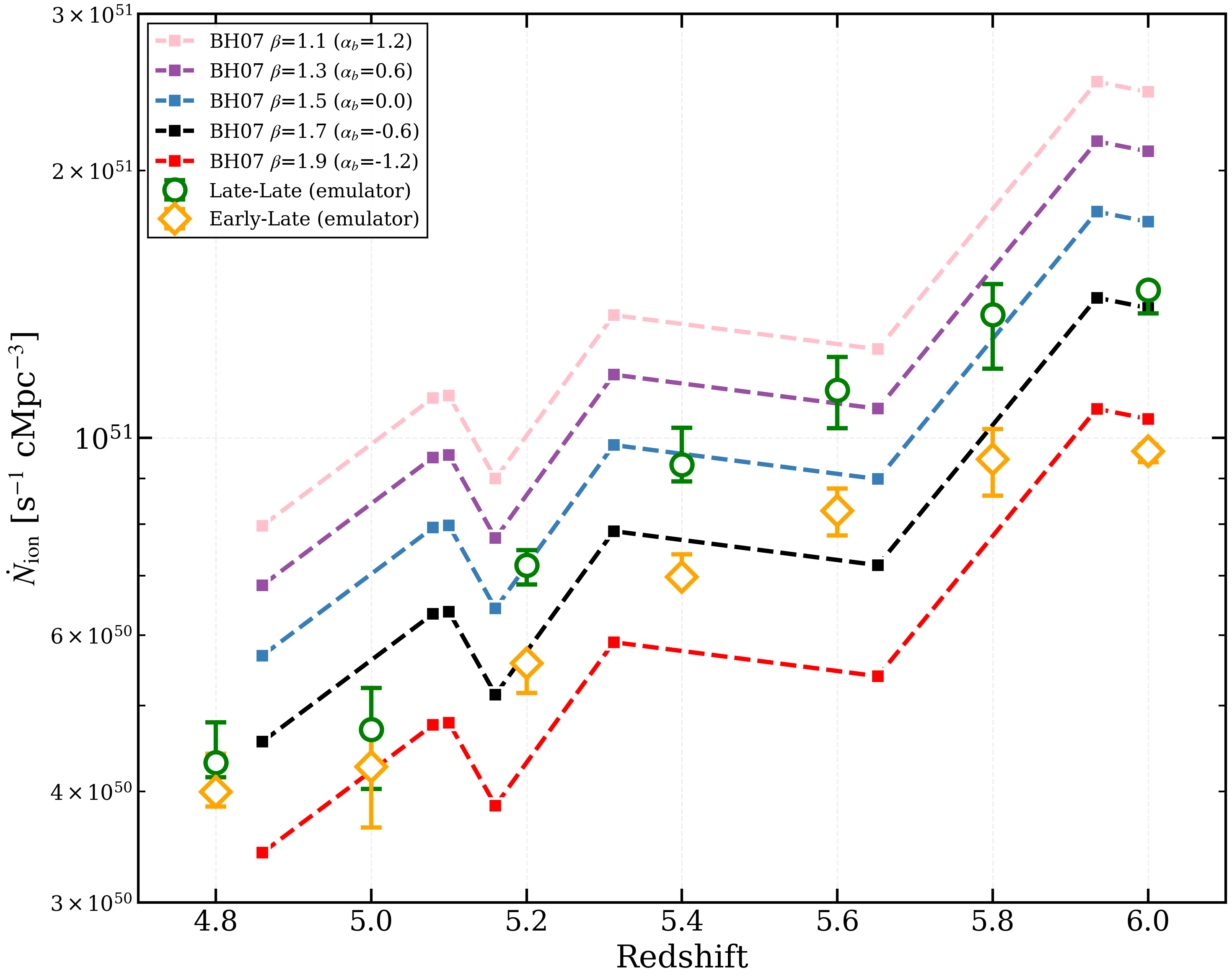}
    \caption{Comparison of ionizing emissivity $\dot{N}_{\rm ion}$ calculated using \cite{2007MNRAS.382..325B} at $z = 4.8$-$6.0$ for different values of the power-law index and direct calculation using our emulator for late-late (green) and early-late (yellow) reionization histories. Green open circles and yellow diamonds with error bars show our emulator-based spectral integration method with $\alpha_s = 1.5$, propagating uncertainties from observed photoionization rates. Colored squares connected by dashed lines show the common power-law approximation for the HI column density distribution that sets the frequency dependence of the MFP, for five values of the power-law index $\beta$.  The corresponding spectral index of the ionizing background, $\alpha_b$, is also labeled.}
    \label{fig:ndot_comparison}
\end{figure}

Figure~\ref{fig:ndot_comparison} compares the two methods at $z = 4.8$-$6.0$. Our spectral integration method (open circles with error bars) propagates uncertainties from the observed photoionization rates, while the power-law approximation (dashed lines) uses the measured mean free path values. Studies typically assume $\beta \approx 1.2$-$1.3$ for the column density distribution slope \cite{2007MNRAS.382..325B, Becker2013}, with values as steep as $\beta \approx 1.9$ sometimes adopted  \cite{Cain2025}. Our emulator-based calculation is comparable to the power-law method with $\beta \approx 1.3$-$1.5$ when $z > 5$ for the late-late reionization model (in green) and 1.7-1.5 for early-late (in yellow) reionization model.  Note that this comparison is not a direct measurement of $\beta$ as it also depends on how closely our model measures MFP values.  Still, the consistency with the emissivity histories inferred assuming lower $\beta$ values may be due to the additional opacity from photoevaporating structures in our late-late and early-late reionization model, which leads to a bump in $d\mathcal{N}/dN_{\rm HI}$ at super Lyman-limit column densities \cite{Nasir2021}.  These structures are eventually evaporated, leading to our emissivity predictions being consistent with a steeper slope at $z<5$.

The constant-$\beta$ models that use the direct MFP measurements (dashed curves) exhibit more jagged evolution than our emulator-based results.  This is mainly because the emulator predicts a smoother MFP evolution that does not go through the center of every measurement (see Figure~\ref{fig:gamma-reionization}).  Interestingly, our emulator still predicts a decline in $\dot{N}_{\rm ion}$ by a factor of $2-3$ (depending on reionization history), just like the constant-$\beta$ models. A decline (or at least flat evolution) in $\dot{N}_{\rm ion}$ below $z = 6$ has been inferred in both direct measurements~\cite{Gaikwad2023,Bosman2024,Cain2025} and is needed in simulations calibrated to agree with quasar-based observables~\cite{Keating2019,Cain2021,2021MNRAS.507.6108O,Gaikwad2023,Cain2024,Asthana2024}.  This is counter-intuitive because the luminosity density of UV sources rises steadily during the same period.  

One possible explanation for this behavior in the measured $\dot{N}_{\rm ion}$ had been the assumption of a constant $\beta$ -- namely, if $\beta$ was decreasing rapidly from $z = 6$ to $5$, the measured evolution in $\dot{N}_{\rm ion}$ would be flatter.  However, our results suggest this is not the case.  We find that the slope of the MFP dependence on frequency evolves from $\approx 1.5$ to $0.9$ between $z = 6$ and $5$ in our emulator models\footnote{$\approx 1.7$ to $1.1$ without using Gaussian Quadrature, showing that capturing density fluctuations on the box scale is somewhat important}.  Since this slope is $\approx 3(\beta-1)$ this translates to an evolution in $\beta$ from $\approx 1.5$ to $1.3$ over this period.  By comparing the blue-dashed and magenta-dashed curves in Figure~\ref{fig:ndot_comparison}, we see that this much evolution in $\beta$ is only enough to explain a $15-20\%$ drop in $\dot{N}_{\rm ion}$, much less than the factor of $2-3$ we infer. It follows that $\beta$ evolution cannot fully explain this behavior.

\section{Data and Code Availability}
The trained emulator, training data, and analysis scripts used in this work are publicly available at \url{https://github.com/htohfa/mean-free-path-emulator}. A frozen version associated with this publication is archived on Zenodo at \href{https://doi.org/10.5281/zenodo.20252450}{10.5281/zenodo.20252450}.

\section{Conclusions}

We developed an emulator for the mean free path of ionizing photons during the epoch of reionization, trained directly on high-resolution radiation-hydrodynamic simulations. Our approach bypasses traditional parameterizations by predicting MFP directly from the physical parameters $z$, $z_{\rm re}$, $\Gamma_{-12}$, and $\delta/\sigma$.

Our residual neural network architecture achieves a median relative error of 1.3\% on held-out test data, with this accuracy maintained across nearly four orders of magnitude in MFP. As a validation test, we ran an additional simulation at $z_{\rm re} = 6.5$, $\Gamma_{-12} = 0.3$ not included in our training suite (and, indeed, is midway in  $z_{\rm re}$ between existing simulations), which the emulator predicts with 1.7\% error at 13.6 eV and 1-3\% errors across all photon energies. The emulator runs in milliseconds, enabling comprehensive parameter studies that would require hundreds of millions of CPU hours using traditional simulations. While our emulator has only learned the MFP for constant $\Gamma_{-12}$ histories, we showed that this results in a negligible bias in our inferences.

We applied the emulator to constrain reionization parameters from observed MFP measurements at $z > 5.0$. Assuming instantaneous reionization and fixed $\Gamma_{-12}$, we grid search over 40,000 parameter combinations and found a best-fit $\Gamma_{-12} = 0.31 \pm 0.22$ s$^{-1}$ and $z_{\rm re} = 5.98 \pm 0.32$. We then extended our analysis to incorporate extended reionization histories by integrating MFP predictions over the distribution of redshifts that IGM gas is reionized. We first considered three reionization histories:  an early-early (starting at $z \sim 12$ and ending around $z \sim 8$), an early-late (starting at $z \sim 12$ and ending around $z \sim 6$), and a late-late reionization (starting at $z \sim 9$ and ending around $z \sim 6$). When using measured $\Gamma_{-12}$ values from observations, the latter two models provide reasonable agreement with MFP measurements at $z = 4.5$-$6$, but the early-early scenario consistently overpredicts the MFP by factors of 2-3. We infer that the measurements prefer a later reionization. We quantified the systematic uncertainty from $\Gamma_{-12}$ evolution and find it introduces up to 20\% uncertainty in MFP predictions, which would shift inferred $z_{\rm re}$ modestly but is insufficient to reconcile the early-early scenario with observations (Appendix~
\ref{sec:evolving_gamma}).

We fit a parametric tanh model directly to MFP observations, finding best-fit parameters $z_{\rm re} = 6.58$ and $\Delta z = 1.63$. Our MFP-based constraints overlap with Planck+kSZ CMB constraints at the 1$\sigma$ level ($z_{\rm re} = 7.8 \pm 0.9$, $\Delta z < 2.8$), demonstrating consistency between these independent probes. The best-fit ionization history implies a neutral fraction of $\approx 30\%$ at $z = 6$, higher than some observational inferences but consistent.  Indeed, in all our models that are consistent at $1\,\sigma$ with the measurements, $10\%$ neutral fractions persist at $z = 6$.

Using our emulator to calculate the ionizing emissivity $\dot{N}_{\rm ion}$ from observed photoionization rates and mean free paths, we find a factor of $2$-$3$ decline between $z=6$ and $z=4.8$, in agreement with previous studies. Our approach avoids assuming power-law scalings for the opacity, instead we directly compute the wavelength-dependent MFP. We find the MFP frequency dependence evolves from slope $\approx 1.5$ to $0.9$ (corresponding to the HI column density distribution slope $\beta$ evolving from $\approx 1.5$ to $1.3$). This $\beta$-evolution accounts for only $15$ -$20\%$ of the $\dot{N}_{\rm ion}$ decline, indicating that changes in the column density distribution cannot explain the emissivity evolution.  Thus, the fast decline in the ionizing emissivity must owe to evolution in the ionizing sources.

Our framework provides a physically motivated and computationally efficient method for predicting the IGM opacity. The emulator can be integrated into large-scale reionization simulations as a subgrid prescription for IGM opacity, bridging the gap between kiloparsec-scale physics captured in our high-resolution simulations and the megaparsec scales needed to model reionization's large-scale morphology. While subgrid models like FlexRT \cite{Cain2021} have demonstrated this approach in principle, our simulation suite combined with the emulator architecture can replace less accurate interpolation schemes currently employed in such models, providing more physically motivated treatment of small-scale opacity fluctuations informed by high-resolution radiative transfer simulations.  One extension that would benefit such studies if they aim for much better than $20\%$ accuracy is to learn how $\Gamma_{-12}$ evolution affects the MFP since this is a smaller effect, we suspect it can be added to the emulator. Our emulator has immediate applications for interpreting observations from JWST \cite{2025ApJ...980...83C} and upcoming extremely large telescopes, including constraining the escape fraction of ionizing photons from early galaxies (which can use the emissivity estimates inferred in this study) and improving theoretical predictions of IGM evolution during and after reionization.

\section{Acknowledgment}
HT and MM acknowledge support from NSF grant AST-2007012 and NASA grant 80NSSC24K1220. AD acknowledges support from NSF grant AST-2045600. We also thank Dr. Hy Trac for allowing us to use his RadHydro code for the simulations.  HT also thanks her Qual reviewers for their helpful comments on the draft.

\appendix

\section{Evolving $\Gamma_{-12}$}
\label{sec:evolving_gamma}
Our fiducial simulations each assume a constant photoionization rate, but in reality $\Gamma_{-12}$ evolves with redshift. This evolution could affect the MFP because the structure of the IGM depends on the photoionization rate history as it takes time for self-shielded gas to photoionize when $\Gamma_{-12}$ goes up and the hydrodynamic response depends on $\Gamma_{-12}$ \cite{Rahmati2013}. However, there are also reasons to think that the evolution of $\Gamma_{-12}$ is a secondary effect on the MFP as the photoheating of optically thin gas does not depend on the amplitude of the ionizing background, and the current ionization state is most sensitive to the current value of $\Gamma_{-12}$. To quantify the potential importance of $\Gamma_{-12}$ evolution, we ran additional simulations with time-evolving $\Gamma_{-12}$ and compared them to our emulator predictions, which use only the instantaneous value of $\Gamma_{-12}$, where over $z=4.5-6$ it evolves by less than a factor of 1.2.

\begin{figure*}
\centering
\includegraphics[width=\textwidth]{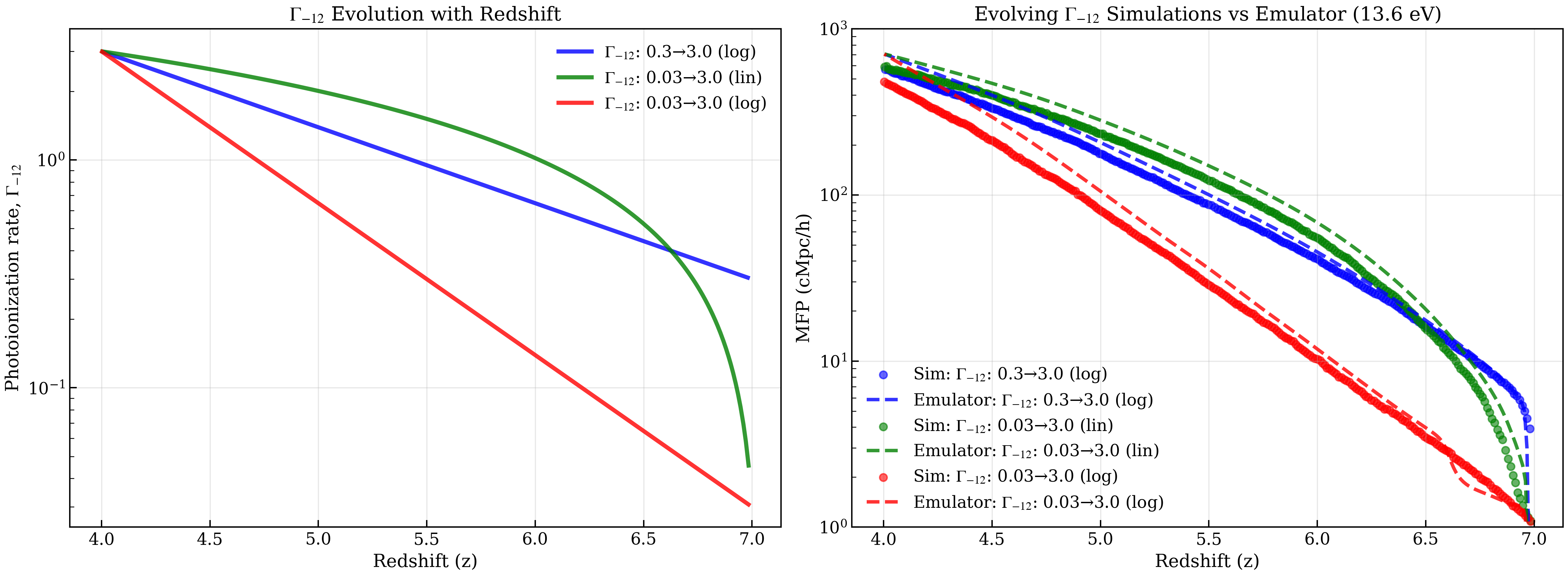}
\caption{Effect of time-evolving photoionization rates on MFP predictions. Left: Three time-evolving $\Gamma_{-12}$ histories as a function of redshift are shown: $\Gamma_{-12} = 0.3 \to 3.0$ evolving linearly in $\log_{10}(z)$ (blue), $\Gamma_{-12} = 0.03 \to 3.0$ evolving linearly in $z$ (green), and $\Gamma_{-12} = 0.03 \to 3.0$ evolving linearly in $\log_{10}(z)$ (red). Right: Comparison of MFP from simulations with time-evolving $\Gamma_{-12}$ (points) to predictions from our emulator using only the instantaneous $\Gamma_{-12}$ value (dashed lines). The emulator tracks the simulations to within $\sim 20\%$, with the largest deviations at intermediate redshifts where cumulative differences in ionization history most affect the gas structure. All simulations assume $z_{\rm re} = 7$.}
\label{fig:evolving_gamma}
\end{figure*}

We consider three scenarios spanning a range of evolutionary histories: (1) $\Gamma_{-12}$ evolving from $0.3$ to $3.0$ linearly in $\log_{10}(z)$, representing a gradual build-up; (2) $\Gamma_{-12}$ evolving from $0.03$ to $3.0$ linearly in $z$, representing rapid early growth followed by slower evolution; and (3) $\Gamma_{-12}$ evolving from $0.03$ to $3.0$ linearly in $\log_{10}(z)$, representing gradual evolution over a larger dynamic range. In all three scenarios, $\Gamma_{-12}$ decreases with increasing redshift, as is expected at redshifts near and during reionization. Note that all of these scenarios represent more extreme evolution at $z<6$ than is observed in the mean $\Gamma_{-12}$~\cite{2018MNRAS.473..560D,Bosman2021,Gaikwad2023}.

Figure~\ref{fig:evolving_gamma} (right panel) compares the MFP from these evolving-$\Gamma_{-12}$ simulations (dots) to predictions from our emulator using only the instantaneous $\Gamma_{-12}$ value at each redshift (dashed lines). The emulator generally tracks the simulations well, with typical differences of $\lesssim 20\%$. These differences are still smaller than the current measurement uncertainties on $\lambda_{\rm mfp}$ at $z > 5$, suggesting that our emulator predictions (which ignore this effect) remain reliable for interpreting existing observations.

\begin{figure*}
\centering
\includegraphics[width=\textwidth]{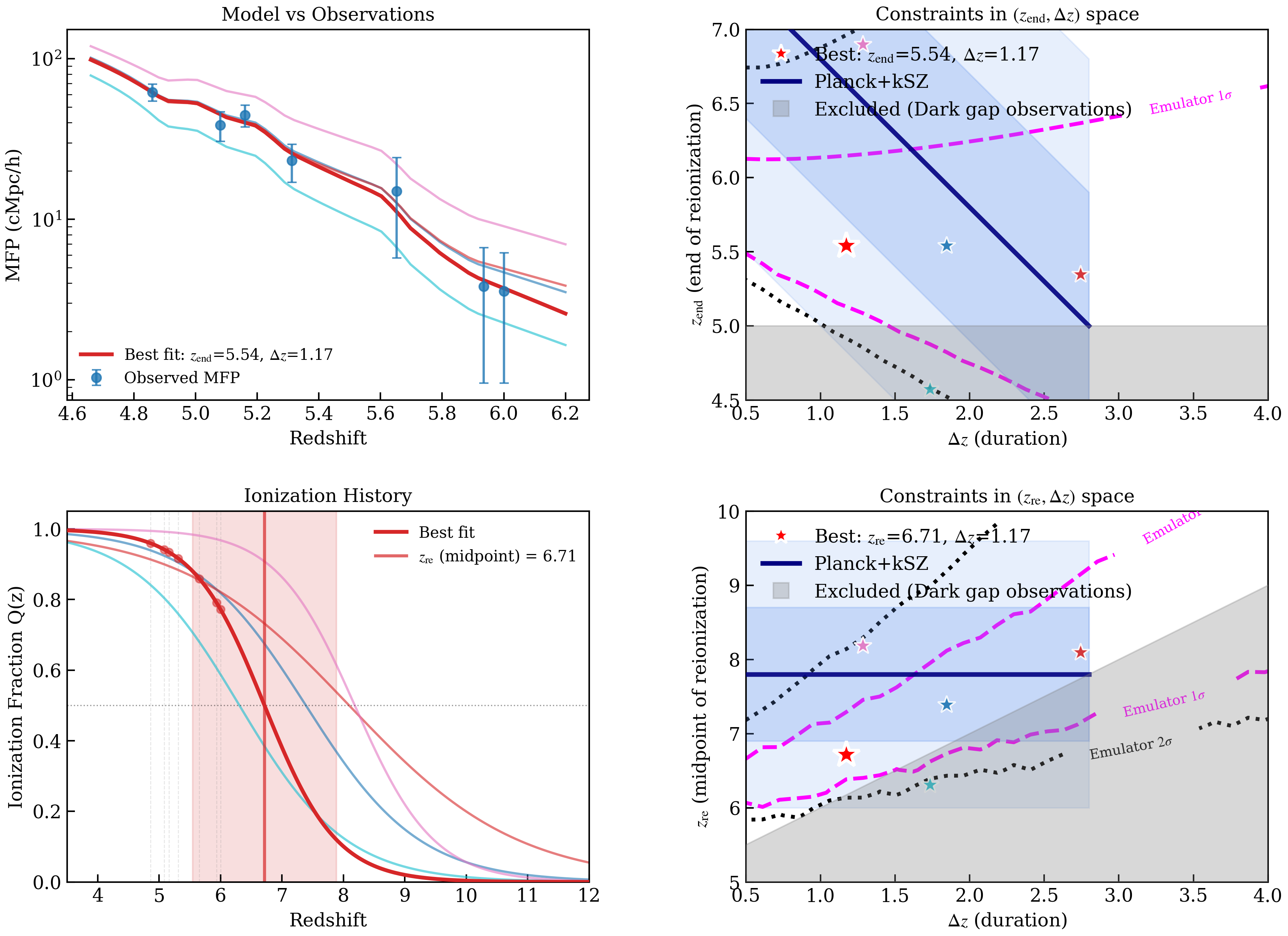}
\caption{Constraints on ionization history when adopting the maximum potential correction for evolution in $\Gamma_{-12}$ (which was not included in the emulator). The layout mirrors Figure~5, but with a 20\% reduction applied to the emulator MFP to account for the maximum potential bias from neglecting $\Gamma_{-12}$ evolution. Top panels show the $(z_{\rm end}, \Delta z)$ space. %The corrected best fit is $(z_{\rm end}, \Delta z) = (5.54, 1.17)$, corresponding to $z_{\rm re} = 6.71$. 
Bottom panels show the ionization history and constraints in $(z_{\rm re}, \Delta z)$ space. %The corrected constraints show better agreement with dark gap observations ($z_{\rm end} > 5$) while maintaining consistency with Planck+kSZ at the $1\sigma$ level.
As described in the text, even adopting this maximum correction does not change our qualitative conclusions.
}
\label{fig:corrected_constraints}
\end{figure*}

To understand how this systematic uncertainty affects our main results, we repeated the tanh parametric fit from Section 3.4 but now applying a 20\% correction to account for the potential bias from neglecting $\Gamma_{-12}$ evolution.  Note that this 20\% correction should overestimate the effect of $\Gamma_{-12}$ evolution given that the observed $z=5-6$ evolution in  $\Gamma_{-12}$ is flatter than the extreme simulations that yield this percent. Figure~\ref{fig:corrected_constraints} shows how the constraints change. The best-fit parameters shift from $(z_{\rm end}, \Delta z) = (4.95, 1.63)$ to $(z_{\rm end}, \Delta z) = (5.54, 1.17)$, with the corresponding midpoint moving from $z_{\rm re} = 6.58$ to $z_{\rm re} = 6.71$.  The 1$\sigma$ contours for  $z_{\rm re}$ and $z_{\rm end}$ shift by similar amount. %The corrected constraints show even better agreement with the dark gap observations, which require $z_{\rm end} > 5$.
%Our best fit now sits comfortably above this threshold rather than in mild tension with it. The consistency with Planck+kSZ constraints remains at the $1\sigma$ level.

This correction also affects the patchy reionization models from Section~3.3. The early-early reionization model, which overpredicted the observed MFP by factors of $\sim$ 2-3, now shows discrepancies of factors $\sim$ 1.5 - 2.5 when we account for the 20\% systematic. This reduces the tension but the early-early scenario remains disfavored at roughly the $2\sigma$ level. The late-late and early-late models continue to agree well with observations, as they were already consistent within the measurement uncertainties. So  $\Gamma_{-12}$ evolution does shift our constraints toward slightly earlier reionization, even in this extreme case of 20\% the main conclusions remain robust. The data still strongly prefer extended reionization histories with substantial neutral fractions persisting to $z \lesssim 6$ over scenarios where reionization completed early at $z > 8$. The shift in best-fit parameters is comparable to our quoted statistical uncertainties. % and crucially, the corrected constraints actually improve the agreement with complementary dark gap observations while maintaining consistency with CMB constraints.

\section{Including correlations between variables}
\label{ap:correlations}
Our approach ignored correlations between reionization redshift, the box-scale linear overdensity, and the ionizing background.  We do not expect these correlations to be important for our results as $\Gamma_{-12}$ and $z_{\rm re}$ tend not to correlate strongly with $ \delta$, as the box is much smaller than the mean free path (especially at the lower redshifts we consider) and the sizes of bubbles during most of reionization. However, here we comment on how these correlations could be included with our approach.

 Our equations generalize to the case of including the full probability distribution of $\Gamma_{-12}$, $z_{\rm re}$ and the box-scale overdensity as
\begin{equation}
\langle  \kappa \rangle = \int dz_{\mathrm{re}}\; d\Gamma_{-12} \; d  \delta \, P(z_{\mathrm{re}}, \Gamma_{-12}, \delta) \, \kappa(z_{\mathrm{re}}, \Gamma_{-12}, \delta),
\label{eq:cddf_averagefull}
\end{equation}
where $ \delta$ is the present-day linear overdensity in our 2 cMpc$/h$ boxes. Here $\kappa(z_{\mathrm{re}}, \Gamma_{-12}, \delta)$ is the inverse of the MFP that the emulator outputs.

Let us briefly comment on how one might calculate $ P(z_{\mathrm{re}}, \Gamma_{-12}, \delta) $. One approach would be to decompose the joint probability distribution as
\begin{equation}
P(z_{\mathrm{re}}, \Gamma_{-12}, \delta) = P(z_{\mathrm{re}} | \delta, \Gamma_{-12}) ~P(\Gamma_{-12}, \delta).
\label{eqn:Psimp}
\end{equation}
One expects $ P(\Gamma_{-12}, \delta)$ to be a Gaussian where the variance is largely set by windowing the matter power spectrum, where different window functions affect $\Gamma_{-12}$  (namely, $e^{-x/\lambda_{\rm mfp}}/x^2$) and $\delta$ (the box shape).  $P(z_{\mathrm{re}} | \delta, \Gamma_{-12})$ is more complicated.  One could use semi-numerical models or simulations to compute this.  Additionally, the highly-successful excursion set formalism for reionization allows an analytic calculation of  $P(z_{\mathrm{re}} | \Gamma_{-12})$ if the barrier is approximated as linear and $\Gamma_{-12}$ is driven by the MFP \cite[see their appendix]{2005ApJ...630..643M} and we expect $P(z_{\mathrm{re}} | \delta, \Gamma_{-12}) \approx P(z_{\mathrm{re}} | \Gamma_{-12})$ for our $2~$Mpc$/h$ boxes since the MFP at the redshifts of interest ($z=5-6$) is much closer to the bubble scale.

\end{document}